\documentstyle[aps,epsfig,prl,floats]{revtex}
\newcommand{\be}{\begin{equation}}
\newcommand{\ee}{\end{equation}}
\newcommand{\bea}{\begin{eqnarray}}
\newcommand{\eea}{\end{eqnarray}}

\begin{document}
\draft
\title{Magnetic Semiconductors are Frustrated Ferromagnets}
\author{Gergely Zar\'and,$^{1,2,3}$ and Boldizs\'ar Jank\'o $^{1,4}$}

\address{
$^1$Materials Science Division, Argonne National Laboratory, 9700
South Cass Avenue, Argonne IL, 60429\\ $^2$Lyman Physics
Laboratory, Harvard University, Cambridge MA 02145 \\ $^3$Research
Group of the Hungarian Academy of Sciences, Institute of Physics,
Technical University Budapest, H-1521 Hungary \\ $^4$ Department
of Physics, University of Notre Dame, Notre Dame IN 46617}
\twocolumn[\hsize\textwidth\columnwidth\hsize\csname
@twocolumnfalse\endcsname

\date{\today}

\maketitle

\begin{abstract}
Starting from  microscopic and symmetry considerations, we derive
the Hamiltonian  describing the exchange interaction between the
localized Mn spins and the valence band holes in $Ga_{1-x}Mn_x
As$. We find that due to the strong spin-orbit coupling in the
valence band, this exchange interaction has a rather complex
structure and generates a highly anisotropic effective interaction
between the Mn spins. The corresponding ground state has a finite
ferromagnetic magnetization but is intrinsically {\em
spin-disordered} even at zero temperature.
\end{abstract}
 ] \narrowtext

Semiconductors resisted for decades the concentrated efforts of a
large community of researchers who wished to turn them into
magnets. The benefits would be far reaching: spin could eventually
be used to carry information in electronic devices. Unfortunately,
any conventional method to produce semiconductor-magnet alloys
failed repeatedly: magnetic materials were insoluble in most
semiconductors. Not long ago \cite{jacek}, several classes of
semiconductor materials finally gave in to a most powerful weapon
against insolubility: the molecular-beam epitaxy (MBE) machine. In
$Ga_{1-x}Mn_x As$, for example, ferromagnetic transition
temperatures as high as  $T_c \sim 110 K$ have been reported
\cite{ohno}. Here we demonstrate, that the victory above the
natural tendency of the semiconductor not to mix with magnetic
materials came with a considerable price. The hostility of the
semiconductor host against magnetic dopants did not simply
disappear, but is ever more present in the {\em highly frustrated}
magnetic correlations that remain in these systems down to the
lowest temperatures. We find, that magnetic semiconductors are by
no means ordinary ferromagnets, but that the intrinsic frustration
pushes these systems into the regime of strongly spin disordered
ferromagnets which also exhibit several features reminiscent of
spin glasses.

We start our analysis by constructing the Hamiltonian that
describes the exchange  interaction between a single magnetic ion
and the valence band holes. In order to be specific, we will
discuss in detail the important case of $\rm Ga_{1-x} Mn_x As$.
However, our results are rather general, as they qualitatively
apply to all those magnetically doped II-VI and  III-V compounds
(and in general to any material with magnetic impurities) which
show strong valence band spin-orbit interaction.

$\rm GaAs$ is a direct gap semiconductor with a valence band
maximum at the center of the first Brillouin zone  called $\Gamma$
point \cite{GaAs}. The top of the valence band is formed by two
two-fold degenerate bands of p-character.  These two valence bands
become degenerate in $\Gamma$. The orbitals involved in these
bands form here a four-dimensional $\Gamma_8$ irreducible
representation. This four-fold degeneracy is due to the strong
spin-orbit interaction that couples the $l=1$ angular momentum of
the p-orbitals to the electron spin ($s=1/2$), thereby producing
an effective total spin $J=l+s = 3/2$ for the valence holes. Since
the third p-band with $J= l-s=1/2$ is separated from the two by a
large spin-orbit splitting, $\Delta_{so}\approx 340\, {\rm meV}$,
for small hole concentrations it is reasonable to describe the
valence band in terms of a two-band model \cite{GaAs,kohn}:
\begin{equation}
H_0 = \gamma_1{p^2 \over 2 m }  - {1\over m} \bigl(
\gamma_2 \sum_\alpha J_{\alpha\alpha} p_{\alpha\alpha}
+ \gamma_3 \sum_{\alpha\ne \beta}  J_{\alpha\beta} p_{\alpha\beta}\bigr)\;,
\label{eq:h0}
\end{equation}
where $m$ is the electron mass and the $\gamma_i$'s are the
so-called Luttinger parameters \cite{GaAs}. The tensor operators
$J_{\alpha\beta}$ and $p_{\alpha\beta}$  ($\alpha,\beta = x,y,x$)
are defined as $Q_{\alpha\beta} = {1 \over 2} (Q_{\alpha}
Q_{\beta} + Q_{\beta} Q_{\alpha}) - {1\over 3}
\delta_{\alpha\beta} {\rm Tr\{ Q_{\alpha\beta}\}}$,  with $Q  = p$
and $J$, referring to the momentum of the  electrons and their
$J=3/2$ effective spin. In the above equation the last two terms,
proportional to $\gamma_2$ and $\gamma_3$, describe the coupling
between the effective spin of the valence hole and its orbital
motion due to spin-orbit interaction. These terms will lift the
four-fold degeneracy for non-zero momenta.

In $\rm Ga_{1-x} Mn_x As$ the Mn ion is believed to  be in the
${\rm Mn}^{2-}$ configuration, corresponding to a half-filled
d-shell with a total spin $S=5/2$ \cite{mnionization}. The general
form of the interaction between the $S=5/2$ Mn spin and the $J=
3/2$-pseudospin  holes depends on the momentum $\bf k$ and $\bf
k'$ of the incoming and outgoing holes. However, close to the
$\Gamma$ point this momentum dependence is weak and we can
approximate the coupling constants by their ${\bf k},\; {\bf
k}'\to 0$ value at the $\Gamma$ point.  We now proceed to
construct a microscopic Anderson-type \cite{anderson} crystal
field model that explicitly takes into account both the local
crystal field symmetry around the Mn impurity and the strong
Coulomb and Hund couplings. We find (G. Z. and B. J., in
preparation) that the dominant part of the interaction has the
following simple form:
\begin{equation}
H_{\rm int}({\bf R})  = G \;
{\bf S} \cdot {\bf J}({\bf R})
\label{eq:coupling}
\end{equation}
with $G$ the exchange coupling, and ${\bf J}({\bf R})$ the spin
density of the holes at the  position ${\bf R}$  of the Mn ion.
Notice that $J$  in Eq.~(\ref{eq:coupling}) denotes the {\em
total} $J=3/2$ spin of the conduction electrons. The above
interaction Hamiltonian can also be established using purely
symmetry considerations. In general, more complicated couplings of
the from $\sim G' \sum_{\alpha}  J_{\alpha\alpha}
S_{\alpha\alpha}$, $\sim G'' \sum_{\alpha\ne \beta}
J_{\alpha\beta} S_{\alpha\beta}$, etc. are also allowed by the
local symmetry $T_d$ of the Mn ion. However, the magnitude of
these couplings turns out to be negligible compared to $G$, due to
the  relatively weak spin-orbit interaction on the Mn ion compared
to the crystal field splitting of the d-levels.

Equations ~(\ref{eq:h0}) and (\ref{eq:coupling}) constitute the
fundamental equations that describe the intricate interplay
between the spin-orbit  interaction in the valence band and the
local moments. Although the model above can be refined to
incorporate the third valence band, it already captures the most
important features necessary to understand the properties of the
ferromagnetic state and related phenomena in magnetic
semiconductors. One of the key differences between this, and
earlier models\cite{bhatt,macdonald} consists in that we now take
into account the spin 3/2 character of the valence holes. This
difference, as we show below, turns out to be crucial for
revealing the true ground state of the system.

In order to analyze the physical content of
Eqs.~(\ref{eq:coupling}) and (\ref{eq:h0}) we first  determine the
effective interaction between two Mn ions at positions  ${\bf
R}_1$ and ${\bf R}_2$, following  the
Ruderman-Kittel-Kasuda-Yoshida (RKKY) procedure \cite{Kittel}.
While the procedure itself is quite straightforward, an explicit
evaluation of a more general type of interaction deduced in this
way is extremely difficult even numerically. Fortunately, for the
case of GaAs host one can make substantial progress by analytical
calculations, provided that we rewrite Eq.~(\ref{eq:h0}) as
\begin{equation}
H_0 = \gamma_1 \bigl({p^2 \over 2 m } - \nu \sum_{\alpha,\beta}
J_{\alpha\beta} p_{\alpha\beta} + \delta H^{(4)} \bigr)\;.
\label{eq:h01}
\end{equation}
The first two parts of the Hamiltonian are rotationally invariant,
$\nu = (6\gamma_3 + 4\gamma_2)/5 \gamma_1 \approx 0.77$, and the
octupolar term $H^{(4)}$ can be shown to  represent only a small
correction \cite{shellmodel}. Therefore, in  leading order we can
set $\delta=0$ and consider only the first two, spherically
symmetric terms in Eq.~(\ref{eq:h01}), which we will denote as
$H_{\rm sp}$.

To diagonalize $H_{\rm sp}$ we choose the spin quantization axis
to be in the $\hat z$-direction. In this basis the energy of plane
waves propagating along the $\hat z$-direction is $\epsilon_\mu(k=
k_z)= {k^2 / 2 m_\mu}$ with $m_\mu = m_h = {m / \gamma_1 (1-\nu)}
\approx 0.5\; m$  and $m_\mu = m_l = {m / \gamma_1 (1 + \nu)}
\approx 0.07\; m $ the heavy and light hole masses for $\mu = \pm
3/2$ and $\mu = \pm 1/2$, respectively. Eigenstates of $H_{\rm
sp}$  propagating in other directions can then be constructed by
simple rotations.  The eigenstates of $H_{\rm sp}$ are {\em
chiral} in nature: The spin of the heavy holes is quantized along
their propagation direction ${\bf \hat k}$ and takes the values
${\bf J \cdot \hat k} = \pm 3/2$. In this new chiral basis $H_{\rm
sp}$ is given by the following simple form:
\begin{equation}
H_{\rm sp}  = \sum_{\bf k,\mu} {k^2\over 2 m_\mu} {c}^\dagger
_{{\bf k}, \mu}{c}_{{\bf k}, \mu}\;,
\end{equation}
where ${c}^\dagger _{{\bf k}, \mu}$ denotes the creation operator
of a hole. In this basis the unperturbed ground state $|0\rangle$
consists of two Fermi spheres. The sphere with the larger radius
contains heavy holes and includes about 90 \% of the valence band
holes, while a sphere with the shorter radius is generated by the
light holes. The price for diagonalizing $H_{\rm sp}$ is that the
exchange coupling in the new basis becomes strongly momentum
dependent:
\begin{equation}
H_{\rm int}({\bf R})  = {G\over V}  \sum_{{\bf k},   {\bf k}'} \sum_\alpha
S^\alpha
\; c^\dagger_{{\bf k} }  J^\alpha(\hat {\bf k}, \hat {\bf k}') c_{{\bf k}}
e^{-({\bf k} - {\bf k}') {\bf R}}\;.
\label{eq:coupling2}
\end{equation}
Here  $J^\alpha(\hat {\bf k}, \hat {\bf k}')$ denotes the operator
$J^\alpha(\hat {\bf k}, \hat {\bf k}') \equiv D^\dagger({\bf \hat
k })   J^\alpha D({\bf\hat k }')$, and $ D({\bf\hat k })$ is the
spin 3/2 rotation matrix, and $V$ the total volume of the sample.
It is precisely this $\hat {\bf k}$-dependence that generates the
delicate magnetic properties of $\rm Ga_{1-x}Mn_xAs$. The
spherical symmetry  of $H_{\rm sp}$ implies that the interaction
between two impurity spins $S_1$ and $S_1$ at a distance $R =
|{\bf R}_1 - {\bf R}_2|$ is given by
\begin{equation}
H_{\rm eff} =
-
 K_{\rm par}(R)\; {\bf S}^{||}_1\cdot {\bf S}_2^{||}
- K_{\rm perp}(R)\; {\bf S}^{\perp}_1 \cdot {\bf S}_2^{\perp} \;,
\label{eq:Heff}
\end{equation}
where ${\bf S}^{||}$ and ${\bf S}^{\perp}$ denote the spin
components parallel and perpendicular to ${\bf R}_1 - {\bf R}_2$.
The form of  $H_{\rm eff} $ is somewhat similar to that of dipolar
interactions as it shows explicit dependence on the relative
position of the Mn impurities. Indeed, the interaction between two
Mn ions far away from each other is in large part mediated by
holes propagating along the axis $\bf R$ that connects them. Since
the majority of the holes are heavy and their spin is quantized
along the propagation direction, it immediately follows that the
interaction must be different for spin components parallel and
perpendicular to $\bf R$.

It turns out that the structure of the effective interaction
Eq.~(\ref{eq:Heff}) can be calculated analytically, although the
details are rather technical and will be reported elsewhere
\cite{long}. The dominant part of the interaction comes from the
heavy hole sector, since this has a much larger density of states
at the Fermi level than the light hole band. The heavy hole
contribution to $K_{\rm par}$ and $K_{\rm perp}$ can be expressed
as
\begin{equation}
K_{\rm par/perp}(R)  = 2\pi \; \epsilon_F \;g_h^2\;
C_{\rm par/perp}(k_{F,h}R)\;,
\label{eq:Cs}
\end{equation}
where $g_h = G \varrho_h$ is the dimensionless heavy hole exchange
coupling, $\varrho_h$ is the heavy hole density of states at the Fermi energy
$\epsilon_F$,
and $k_{F,h}$ denotes the heavy hole Fermi momentum. The dimensionless
functions $C_{\rm parp/perp}(y)$ are clearly different
(see  Fig.~\ref{fig:rkky}), and in
the  $y \to 0$ limit  are approximately given by
$C_{\rm perp}(y)\approx 1/y$ and $C_{\rm par}(y)\approx 1/2y$.

In $Ga_{1-x}Mn_xAs$ only a small fraction $f$ of the Mn ions gives
a hole into the valence band. Although the exact value of this
fraction is not precisely known, latest experiments suggest
\cite{dietl}, that for x=0.05 Mn concentration (corresponding to
the highest $T_c$) this fraction is about $f=0.2-0.3$  (or
$k_{F,h} \approx 0.14 1/\AA$) and a typical Mn-Mn distance is
approximately $d_{\rm Mn-Mn}\approx 12\AA$. Thus, for typical
nearest-neighbor Mn ions $K_{\rm perp}$ is larger than  $K_{\rm
par}$ and   {\em ferromagnetic}. However, this ferromagnetic
interaction is {\em strongly anisotropic} as it tries to align
nearby pair of Mn spins parallel to the axis connecting them (see
the illustration in Fig.~\ref{fig:rkky}). Since it is clearly
impossible  to simultaneously satisfy each  pair of spins, this
effect induces {\em orientational frustration} and influences the
magnetic properties of $Ga_{1-x}Mn_xAs$ in a fundamental way.
\begin{figure}[tb]
\begin{center}
\epsfxsize7.5cm
\epsfbox{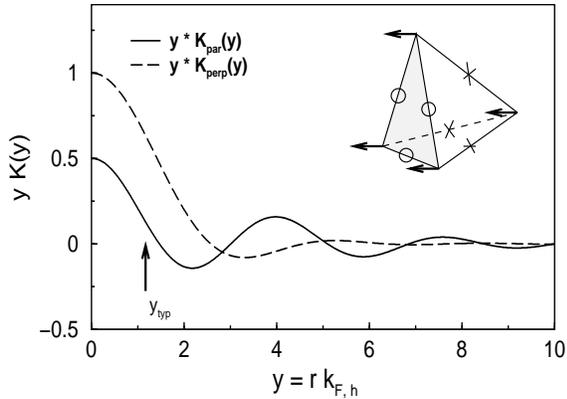}
\end{center}
\vskip0.1cm \caption{\label{fig:rkky} Main figure: Spatial dependence of the
two interaction parameters $C_{\rm perp}$ and $C_{\rm par}$ of
Eq.~(\protect{\ref{eq:Cs}}). Only the contribution of the heavy
hole sector is shown.  The arrow indicates the
typical value of $y$ for $x\approx 0.05$ and $f \approx 0.25$.
Inset: Two  Mn impurities close to each-other tend to be aligned {\em
perpendicular} to the axis connecting them. If we satisfy the bonds
marked by circles the other bonds with crosses remain
unsatisfied, resulting in orientational frustration.}
\end{figure}

We investigated the implications of the anisotropy
\cite{anisotropy} on the magnetic properties of $Ga_{1-x}Mn_xAs$
by performing classical Monte Carlo (MC) simulations using the
effective interaction of  Eq.~(\ref{eq:Heff}). In the
simulations the Mn spins were replaced by classical angular
variables, ${\bf S} \to S \;{\bf \Omega}$.
The $Mn$ ions were randomly distributed on a $N \times
N \times N$ face-centered cubic lattice with a probability
$x=0.05$. $N = L/a$ is the linear extension of the lattice in
units of the conventional lattice constant $a=5.65 \AA$. To take
into account the finite mean free path $l\approx 7\AA$
of the valence holes, we used an exponential cutoff for
the RKKY interaction \cite{ohnox}: $K_{\rm par/perp}(R)
\to K_{\rm par/perp}(R) e^{-R/l}$.

\begin{figure}[tb]
\begin{center}
\epsfxsize7.5cm
\epsfbox{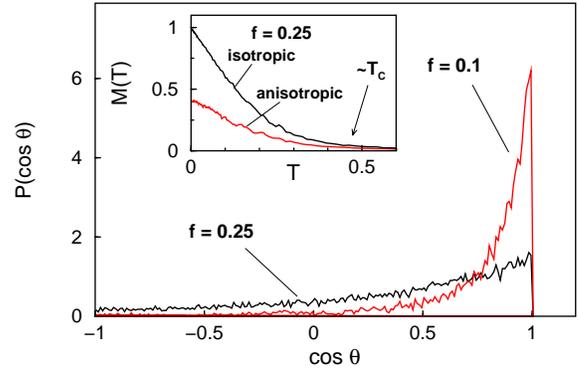}
\end{center}
\vskip0.1cm \caption{\label{fig:distribution} Ground state
probability distribution function of $\cos\theta = \bf \Omega
\cdot n$. The unit vector $\bf n$ gives the direction of the
magnetization in the ground state. In the isotropic approximation,
when  $K_{\rm par} = K_{\rm perp} = (2 K_{\rm par} + K_{\rm
perp})/3$, the ground state is almost fully  polarized, and thus
$P(\cos\theta) =\delta (\cos\theta-1) $. Inset: Magnetization as a
function of temperature for these two cases. Temperature is
measured in units of  $ 2\pi \; \epsilon_F \;g_h^2 S^2$. For the
isotropic interaction the magnetization saturates as $T\to0$. }
\end{figure}

In the inset of Fig.~\ref{fig:distribution} we show the
magnetization $M\equiv|\langle \Omega_i\rangle|$ as a function of
temperature. We find that a spontaneous magnetization develops at
low temperatures \cite{lowtemp}. The transition between the
paramagnetic and magnetic phase takes place rather smoothly, and
then increases approximately linearly with decreasing temperature.
Both properties agree  qualitatively with the experiments
\cite{magnlinear}, and are characteristic to strongly disordered
magnets \cite{bhatt}. The spontaneous magnetization,
however, tends to a value at $T=0$ that is much {\em smaller} then
that of a fully polarized ferromagnet. This reduction is clearly
due to orientational disorder originating in the anisotropy of the
interaction, and has nothing to do with possible antiferromagnetic
couplings due to the RKKY oscillations of  the Mn-Mn interaction.
To demonstrate this, we repeated the simulations  by replacing the
interaction in Eq.~(\ref{eq:Heff}) by its angular average. As
shown in Fig.~\ref{fig:distribution}, the magnetization in this
case {\em saturates} to its maximal value (normalized to unity),
and all the Mn spins are fully polarized.

More information can be obtained about the ground state properties
by measuring the distribution of the product $\cos\theta \equiv
{\bf \Omega}_i {\bf n}$, where  ${\bf n}$ is the a unit vector
parallel to the ground state magnetization. Without the spatial
anisotropy structure discussed here, $P(\cos\theta) =\delta
(\cos\theta-1) $, since the spins are fully aligned. As shown in
Fig.~\ref{fig:distribution}, in the system with the correct
exchange interaction, the quantity $\cos\theta$ has a very broad
but asymmetric distribution. Depending on the actual value of $f$
the distribution has more or less weight in the vicinity of
$\cos\theta =1$: For $f=0.1$ the Mn spins tend to  point
approximately into the direction of the global  magnetization,
however, they deviate  at the average by an angle $\theta \sim 30$
degrees and the magnetization is reduced considerably by $\sim
20\%$. The ground state becomes more spin disordered for larger
carrier fractions: The distribution has less weight at $\cos\theta
= 1$ and  many of the spins are aligned {\em
antiferromagnetically} with respect to the global magnetization,
thus reducing the magnetization by about $\sim 50\%$.

The large reduction in the magnetization we find here has been
observed  experimentally \cite{magnlinear,magnfield}. The measured
in-plane saturation magnetization of GaMnAs is about 50 $\%$ less
than the value that would correspond  to the known Mn
concentration. Also, cooling down the sample in a relatively weak
($B \leq 4 \, T$) external field results in a 20-40$\%$  increase
in the $T=0$ magnetization. Also, the broad linewidth of
ferromagnetic resonance data is consistent with the substantial
intrinsic spin-disorder reported here.\cite{Furdyna2}

The results of our simulations are consistent with a highly spin
disordered ferromagnetic ground state with spin glass-like
behavior. Indeed, we found many metastable, {\em macroscopically}
different local energy minima of the Hamiltonian, extremely  close
to the ground state, a characteristic property of spin glass
systems. The precise nature of the ground state can be determined
experimentally. One of the typical experimental signatures of a
spin glass state would be the history dependence of the high field
magnetization in fields parallel to the film, or the difference
between field cooled and zero field cooled
susceptibilities.
This is a straightforward experiment, which can be performed with
already existing samples and apparatus. However, the outcome would
provide a highly valuable insight into the true order present in
these magnetically ordered semiconducting systems.

The intrinsic spin disorder we described above could be the reason
for various resistance anomalies. Since this intrinsic spin
disorder produces a large spin scattering contribution to the
resistivity, it may provide  an explanation to the anomalous
magnetoresistance of  GaMnAs alloys with smaller $T_c$ at
temperatures $T\ll T_c$. It can also possibly explain why these
matierals exhibit a resistance peak {\em precisely} at the
ferromagnetic phase transition:  Due to the spin disorder present
in the ground state, the magnetic order parameter which appears at
$T_c$ also involves 'disordered' fluctuations corresponding to
{\em finite} momenta. These finite momentum components of the soft
modes may result in a maximum of the resistance at $T_c$. This
should be contrasted to any conventional ferromagnet, which
exhibits predominantly long wavelength soft modes and, therefore
shows no peak in the resistivity at $T_c$ \cite{littlewood}.

In conclusion, we have computed the effective exchange interaction
strong spin-orbit interaction in its valence band. We found that
the resulting exchange interaction inevitably shows a highly
anisotropic structure, which in turn generates a strongly spin
disordered ground state. We have shown that the qualitative
features of the experimentally measured magnetization can be
reproduced by a straightforward simulation of a spin system
governed by the effective exchange interaction we derived. Our
results  suggest that the actual experimental systems are
disordered ferromagnets with features reminiscent of spin glasses,
and proposed an experimental test of such glassy behavior.

We are grateful to Professors Peter B. Littlewood, Allan H.
MacDonald, Peter Schiffer and especially Jacek K. Furdyna for
stimulating discussions. This research has been supported by the
U.S. DOE, Office of Science, the NSF Grants Nos. DMR-9985978 and
DMR97-14725, and Hungarian Grants No. OTKA F030041,  T029813, and
T29236. G. Z. is an E\"otv\"os fellow. 

\begin{references}
\bibitem{lowtemp}
Well below $T_c$, the valence bands become partially
spin-polarized. While the polarization of the carriers could
influence the quantitative value of the magnetization, all the
qualitative aspects of the ordered state are already captured by
the simulation described above.
\end{references}
\end{document}